\documentclass[fleqn,twoside]{elsart4}
\usepackage{graphicx,amssymb}

\begin{document}

\begin{frontmatter}

\title{Magnetic exchange coupling and Curie temperature of Ni$_{(1+x)}$MnSb ( x=0, 0.25, 0.5, 0.75, 1) from first principles}

\author{E.~\c Sa\c s\i o\~glu\corauthref{cor1}} 
\author{, L. M.  Sandratskii, and  P. Bruno}

\address{Max-Planck Institut f\"ur Mikrostrukturphysik,
D-06120 Halle, Germany}

\corauth[cor1]{ersoy@mpi-halle.de}

\begin{abstract}

We  study  the dependence  of magnetic interactions and  Curie
temperature in Ni$_{(1+x)}$MnSb system   on the Ni concentration
within the framework of the density-functional theory. The
calculation of the exchange parameters is based on the super--cell
and frozen-magnon approaches. The Curie temperatures, T$_C$, are
calculated within the random-phase approximation. In agreement
with experiment we obtain decrease of the Curie temperature with
increasing Ni content.

\end{abstract}

\end{frontmatter}

Recently rapid development of spin electronics intensified  the
researches on the  ferromagnetic  materials that are suitable for
the  spin injection  into a semiconductor\cite{ohno}. One of the
promising classes of these materials is the Heusler alloys.
Heusler alloys have been intensively studied theoretically
\cite{groot,galana} and experimentally \cite{exp_1,exp_2} as
possible sources of spin-polarized carriers for spintronics
applications. Among the properties useful for the applications are
high Curie temperature, high electron spin polarization at the
Fermi level and very small lattice mismatch with widely employed
semiconductors (e.g., Ni$_2$MnIn and InAs)\cite{inas_3}. Some of
the Heusler compounds were found to have half-metallic ground
state \cite{groot} which is characterized by a 100\% carrier
spin-polarization. An interesting combination of physical
properties makes Heusler alloys the subject of intensive
experimental and theoretical investigations
\cite{ref_7,ref_9,ref_10}.

In the present work we report the theoretical study of the
exchange interactions and Curie temperature of Ni$_{(1+x)}$MnSb
system as a function of  Ni concentration. In   particular, we
focus on the effect of  local environment of Mn atoms on
half-metallicity and exchange interactions. We  show that,
occupation of vacant  sites in NiMnSb by  Ni immediately  leads
to disappearance of  the half-metallicity and to substantial
decrease of both inter-sublattice (Mn-Ni) and intra-sublattice
(Mn-Mn) exchange interactions.

The calculations are carried out with the augmented spherical
waves (ASW) method \cite{asw} within the generalized gradient
approximation (GGA) \cite{gga} for the  exchange--correlation
potential. In all calculations the experimental values of the
lattice parameters are used \cite{webster}. The radii of all
atomic spheres are chosen equal \cite{radii_exp}.  We use cubic
super cell for $0.25\leq x \leq 0.75$.

We describe the interatomic exchange interactions in terms of the
classical Heisenberg Hamiltonian
\begin{equation}
\label{eq:hamiltonian2}
 H_{eff}=-  \sum_{\mu,\nu}\sum_{\begin {array}{c}
^{{\bf R},{\bf R'}}\\ ^{(\mu{\bf R} \ne \nu{\bf R'})}\\
\end{array}} J_{{\bf R}{\bf R'}}^{\mu\nu}
{\bf s}_{\bf R}^{\mu}{\bf s}_{\bf R'}^{\nu}
\end{equation}
In Eq.(\ref{eq:hamiltonian2}), the  indices  $\mu$ and $\nu$
number different sublattices and ${\bf R}$ and ${\bf R'}$ are the
lattice vectors specifying the atoms within sublattices, ${\bf
s}_{\bf R}^\mu$ is the unit vector pointing in the direction of
the magnetic moment at site $(\mu,{\bf R})$.

We employ the frozen--magnon approach \cite{magnon_2,magnon_3} to
calculate interatomic Heisenberg exchange parameters. The
calculations involve few steps. In the first step, the exchange
parameters between the atoms of a given sublattice $\mu$ are
computed. The calculation is based on the evaluation of the energy
of the frozen--magnon configurations defined by the following
atomic polar and azimuthal angles
\begin{equation}
\theta_{\bf R}^{\mu}=\theta, \:\: \phi_{\bf R}^{\mu}={\bf q \cdot
R}+\phi^{\mu}. \label{eq_magnon}
\end{equation}
The constant phase $\phi^{\mu}$ is always chosen equal to zero.
The magnetic moments of all other sublattices are kept parallel to
the z axis. Within the Heisenberg model~(\ref{eq:hamiltonian2})
the energy of such configuration takes the form
\begin{equation}
\label{eq:e_of_q} E^{\mu\mu}(\theta,{\bf
q})=E_0^{\mu\mu}(\theta)+\sin^{2}\theta J^{\mu\mu}({\bf q})
\end{equation}
where $E_0^{\mu\mu}$ does not depend on {\bf q} and the Fourier transform $J^{\mu\nu}({\bf q})$
is defined by
\begin{equation}
\label{eq:J_q}
J^{\mu\nu}({\bf q})=\sum_{\bf R}
J_{0{\bf R}}^{\mu\nu}\:\exp(i{\bf q\cdot R}).
\end{equation}

In the case of $\nu=\mu$ the sum in Eq. (\ref{eq:J_q}) does not
include ${\bf R}=0$. Calculating $ E^{\mu\mu}(\theta,{\bf q})$ for
a regular ${\bf q}$--mesh in the Brillouin zone of the crystal and
performing back Fourier transformation one gets exchange
parameters $J_{0{\bf R}}^{\mu\mu}$ for sublattice $\mu$.

The  determination  the exchange interactions between the atoms of
two different sublattices $\mu$ and $\nu$  is discussed in Ref.
\cite{intersublattice}.

The Curie  temperature is estimated  within the random phase
approximation (RPA)
\begin{equation}
\label{eq:rpa}
\frac{1}{k_{B}T_{c}^{RPA}}=\frac{6\mu_B}{M}\frac{1}{N}\sum_{q}\frac{1}{\omega({\bf
q })}
\end{equation}
where $ \omega({\bf q})$ is  the   spin--wave dispersion.

We begin  with the discussion of the structural properties. NiMnSb
and Ni$_2$MnSb crystalize in C1$_b$ type and in L2$_1$ type
structures respectively. According to our calculations for the
intermediate compositions ($0<x<1$) lower total energy corresponds
to  the C1$_b$ type structure with  one  Ni sublattice completely
and another Ni sublattice  partially filled. This  result is   in
good agrement with the experiment \cite{webster}.

\begin{figure}[ht]
\begin{center}
\includegraphics[scale=0.35]{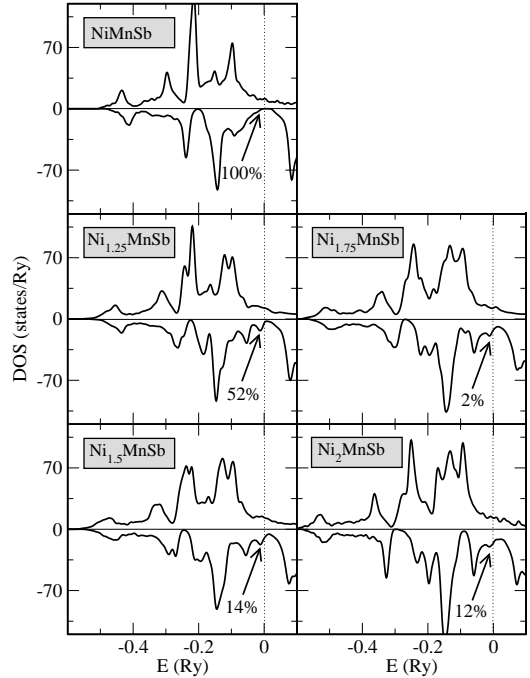}
\end{center}
\caption{
 Spin--projected total density of states  of Ni$_{(1+x)}$MnSb.
The  spin polarization at the  Fermi level is shown. }
\label{total_dos}
\end{figure}

We present  in Fig.\ref{total_dos}  calculated  total density of
states (DOS). NiMnSb is found to have half--metallic ground state
which is characterized by a 100 \% carrier spin-polarization at
the Fermi level. The occupation of vacant sites in NiMnSb by Ni atoms
destroys the  half--metallicity and reduces the spin polarization.
The loss of the half-metallicity with increasing Ni concentration can
be attributed to  the increased Mn--Ni hybridization
\cite{galana}. In table \ref{tab:moments} we present calculated
magnetic moments. The Mn and  Ni$^1$  magnetic moments decreases
with increasing Ni concentration. On the other hand the Ni$^2$
moment increases. Total magnetic  moment slightly decreases.

\begin{table}
\caption{magnetic moments (in $\mu_B$) Ni$_{(1+x)}$MnSb. Ni$^1$
and Ni$^2$ denote first  and second Ni sublattices, respectively.
Different values of magnetic moments in Ni$^1$ column  corresponds
to inequivalent atoms in first Ni sublattice.}
\begin{center}
\begin{tabular}{lcccccccc}
\\ \hline
 \hline

            $ $ & $ $ &  Ni$^1$     & $ $ &   Ni$^2$ & $ $ & Mn &  $ $ &Total\\ \hline
NiMnSb          & $ $ & 0.19        & $ $ &  -    & $ $ &3.86   & $ $ & 4.000 \\
Ni$_{1.25}$MnSb & $ $ & (0.20-0.21) & $ $ & -0.04 & $ $ & 3.82  & $ $ & 3.971 \\
Ni$_{1.5}$MnSb  & $ $ & (0.19-0.21) & $ $ & 0.03  & $ $ & 3.81  & $ $ & 3.976 \\
Ni$_{1.75}$MnSb & $ $ & (0.18-0.19) & $ $ & 0.08  & $ $ & 3.76  & $ $ & 3.956 \\
Ni$_2$MnSb      & $ $ & 0.14        & $ $ & 0.14  & $ $ & 3.695 & $ $ & 3.945 \\
 \hline
  \hline
\end{tabular}
\end{center}
 \label{tab:moments}
\end{table}

\begin{table}
\caption{RPA estimation of the Curie temperature in
Ni$_{(1+x)}$MnSb. Experimental values of the Curie temperatures
are taken  from Ref. \cite{webster}.}
\begin{center}
\begin{tabular}{lcccc}
\\ \hline
 \hline
            $ $ & $ $ &  T$_c$(RPA)     & $ $ &   T$_c$(Exp) \\ \hline
NiMnSb          & $ $ &   830       & $ $ &  730   \\
Ni$_{1.25}$MnSb & $ $ &   550       & $ $ &  570   \\
Ni$_{1.5}$MnSb  & $ $ &   364       & $ $ &  470   \\
Ni$_{1.75}$MnSb & $ $ &   254       & $ $ &  405   \\
Ni$_2$MnSb      & $ $ &   190       & $ $ &  365   \\
 \hline
  \hline
\end{tabular}
\end{center}
 \label{rpa_tc}
\end{table}

\begin{figure}[ht]
\begin{center}
\includegraphics[scale=0.34]{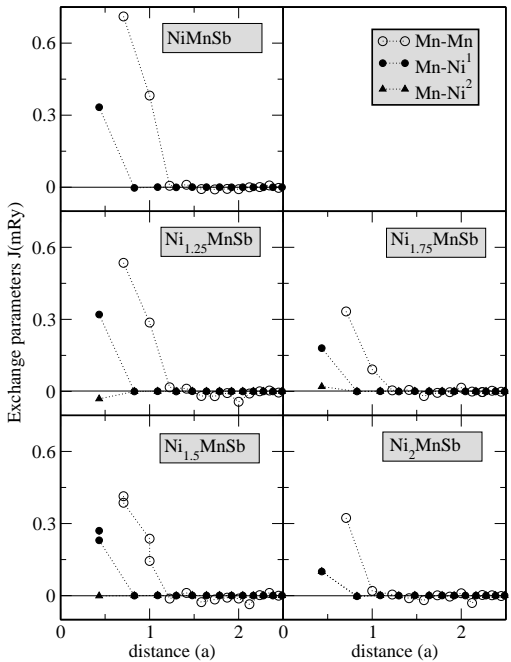}
\end{center}
\caption{ Intra--sublattice (Mn--Mn) and  iter--sublattice
(Mn--Ni) exchange interactions in Ni$_{(1+x)}$MnSb. The distances
are given in the units of the lattice constant. }
 \label{exchange}
\end{figure}

The  calculated Heisenberg exchange  parameters are presented  in
figure \ref{exchange}. The decrease in the  experimental  Curie
temperatures with increasing  Ni concentration is in good agrement
with the calculated parameters. Transition from NiMnSb to
Ni$_2$MnSb leads to  a strong decrease in all exchange parameters
except Mn-Ni$^2$ which increases slightly . Although the symmetry
of the system is reduced at the intermediate compositions, the
symmetry  of the exchange parameters  at small ($x\leq 0.25$) and large ($x\geq
0.75$) values of $x$ is not affected by this
reduction. Only at $x=0.5$ the pattern of exchange
interactions shows anisotropic behavior appeared as a
splitting  in  nearest and next nearest neighborhood exchange
parameters. The  splitting in further nearest neighborhoods can be
neglected.

The interatomic exchange parameters are used to evaluate the Curie
temperature. We take  only the Mn--Mn interactions and neglect
Mn--Ni contribution. As it was shown  in
Ref.\cite{intersublattice} the contribution of the Mn--Ni exchange
interaction to the Curie temperature of Ni$_2$MnSb is much   less
important than Mn--Mn contribution.
The calculated Curie temperatures are presented in table \ref{rpa_tc}. For
low $x$ ($x \leq 0.5$) the values of  T$_C$ are  in good  agrement
with the experiment.  For large $x$, the calculations
underestimate T$_C$.

In conclusion, we  have  systematically  studied the dependence of
magnetic interactions and  Curie temperature in Ni$_{(1+x)}$MnSb
system   on the Ni content within the parameter--free
density functional theory. Our  calculations show that Ni
substitution destroys the half-metallicity and substantially
decreases both inter-sublattice (Mn-Ni) and intra-sublattice
(Mn-Mn) exchange interactions. The Curie temperatures are
calculated within the random phase approximation to
the statistical mechanics of the classical
Heisenberg Hamiltonian. In agreement with experiment we obtain
decrease of the Curie temperature with increasing Ni content.

\end{document}